# Focusing and Diffraction of Light by Periodic Si Micropyramidal Arrays

Grant W. Bidney,[1,2] Amstrong R. Jean,[1] Joshua M. Duran,[2] Gamini Ariyawansa,[2] Igor Anisimov,[2]
Kenneth W. Allen,[3] and Vasily N. Astratov[1,2,*]

[1]Department of Physics and Optical Science, Center for Optoelectronics and Optical Communications,
University of North Carolina at Charlotte, Charlotte, NC 28223-0001, USA
[2]Air Force Research Laboratory, Sensors Directorate, Wright Patterson AFB, OH 45433, USA
[3]Advanced Concepts Laboratory, Georgia Tech Research Institute, Georgia Institute of Technology, Atlanta, GA 30332, USA
*Tel: 1 (704) 687 8131, Fax: 1 (704) 687 8197, E-mail: astratov@uncc.edu

*Abstract*— This research was devoted to modeling of the optical properties of Si micropyramids aimed at designing optimal structures for applications as light concentrators in mid-wave infrared (MWIR) focal place arrays (FPAs). It is shown that completely different optical properties of such structures can be realized using two types of boundary conditions (BCs): i) periodical and ii) perfectly matched layer. The first type (periodical BC) allowed us to describe the Talbot effect under plane wave coherent illumination conditions. This effect was experimentally demonstrated in the proposed structures. The second type (perfectly matched layer BC) allows describing the optical properties of individual micropyramids concentrating or "focusing" light on the photodetector. The optimal geometries of micropyramids required for maximizing the intensity of "photonic nanojets" emerging from their truncated tips are determined.

*Keywords*— infrared photodetectors, light concentrators, dielectric resonance

## I. INTRODUCTION

In the recent few years, we proposed that anisotropic wet etching of Si can be used as a novel way to fabricate light concentrators for mid-wave infrared (MWIR) focal plane arrays (FPAs) [1-3]. Previously, this technology was used by the microelectromechanical (MEMS) community and its optical applications were rather limited. This method enables fast and parallel fabrication large-scale micropyramidal arrays with smooth sidewall surfaces that is attractive for optical applications. In our previous work, however, the analysis of the optical properties was limited to a fixed geometry of microcones with 14 μm larger base and the smaller base varying around 4 μm [3?]. It created a question about the role the micropyramid's geometrical parameters have regarding their optical properties. Since micropyramidal arrays diffract light beams, it creates a broader question about how one can model these "grating" properties. On the other hand, in a final application, each micropyramid focuses light onto its own photodetector and the intensity enhancement factors (IEFs) on the detectors need to be estimated. In this final application, the incident light is typically incoherent, and the role of diffraction effects is reduced – each micropyramid concentrates light onto its own photodetector, however some crosstalk cannot be completely excluded.

In this work, the answers to these questions were obtained by carefully selecting the boundary conditions (BCs) for the problem. This approach allowed the modeling to predict and stimulated the the experimental observation of the Talbot effect in the fabricated structures. It also allowed optimization of the micropyramid's geometry required for achieving maximal IEFs of the "photonic nanojets" produced near the tips of the truncated micropyramids.

## II. SIMULATION AND EXPERIMENTAL RESULTS

### A. Role of the Boundary Conditions (BCs)

Finite-difference time-domain (FDTD) software by Lumerical was used to run the computer simulations. The modeled object was a truncated Si micropyramid with the refractive index $n=3.5$. Since the slope of the sidewall surface was fixed at 54.7° by etching, the variable parameters were the sizes of the micropyramid bases in Fig. 1. The source of plane waves was embedded in a Si wafer.

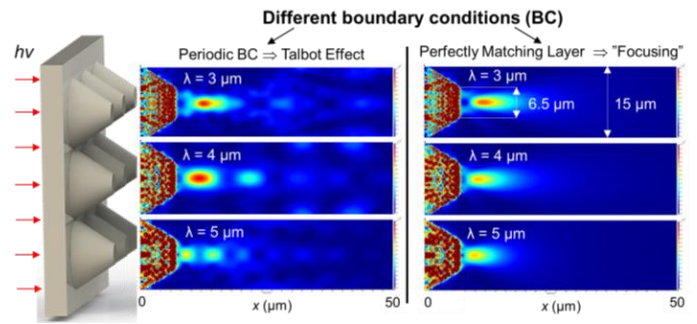

Fig. 1. EM field distributions calculated at normal incidence for truncated Si micropyramids ($n=3.5$) with 15 μm large base and 6.5 μm small base. In the case of periodic BC there are multiple EM peaks due to the Talbot effect. In the case of perfectly matched layer BC, there is a single "photonic nanojet" which appears similar to the focusing of light by a lens.

It was found that the BCs play a significant role in the results of such electromagnetic (EM) modeling. Selection of the periodic BCs means that the calculations effectively represent the collective properties of an infinite periodic array, as illustrated in the left side of Fig. 1. It is seen that the EM field distribution illustrates periodical peaks due to diffraction and interference effects introduced by the micropyramidal array similar to the case of diffraction grating. This phenomenon is called the Talbot effect and it is generally known for periodic structures [5].

On the other hand, selection of the perfectly matched layer BCs means that the EM waves which reach the boundary of the computational area are allowed to escape the area. Thus, this type of BCs describes the behavior of individual micropyramids without considering contributions from neighboring structures due to diffraction and interference effects. As a result, the calculations show a single EM peak, as illustrated in the right part of Fig. 1. By analogy with the case of dielectric microspheres, such EM peaks can be termed "photonic nanojets" [6] and this terminology becomes widely accepted for microscale structures with different shapes. It is seen that the position of this peak depends on the wavelength ($\lambda$), see the results calculated for $\lambda$=3, 4, and 5 µm in the right part of Fig. 1. This case is closer to the practical operation of micropyramids integrated with photodetectors.

### B. Talbot Effect Modeling: Periodic BC

To prove that EM peaks observed using periodic BC are due to the Talbot effect, we studied the dependence of the calculated EM maps on the period of the array ($A$), which is equal to the size of the micropyramid's large base (the neighboring micropyramids are touching). This dependence is illustrated in Fig. 2 for $A$ = 5, 8, and 11 µm. According to the theory of the Talbot effect, the distance between the neighboring EM maxima should be equal to the Talbot length ($x_T$) [5]:

$$x_T = \lambda/[1 - (1 - \lambda^2/A^2)^{1/2}]. \quad (1)$$

It was found that the distance between the neighboring EM peaks in Fig. 2 follows the Talbot length in good agreement with Eq. (1), thus confirming that the peaks are due to the Talbot effect. The fact that, in some cases, the EM field peaks appear with multiple maxima is likely due to the complicated shape of the truncated micropyramids compared to the simplest model consisting of a single-period sinusoidal diffraction grating.

### C. Experimental Observation of the Talbot Effect

Experimentally, the Talbot effect was studied using a setup illustrated in Fig. 3(a). Illumination was provided by a Er:YAG laser at $\lambda$=2.96 µm slightly focused to ~0.5 mm spot size on the micropyramidal array to increase the intensity. It can be viewed as a quasi-plane wave illumination similar to our theoretical model. The transversal intensity distributions at different imaging planes were projected on the MWIR Spiricon beam profiler using a Ge or CaF lens transparent in the MWIR range. Scanning of the 3-D intensity distribution was performed by translating the lens along the optical axis of the system with micrometer precision.

Scanning the imaging plane along the optical axis ($x$) revealed multiple positions where the sharply focused peaks were observable, as illustrated in Fig. 3(b). These results were obtained using a micropyramidal array with 30 µm pitch and 16 µm size smaller base. The brightest image takes place at the focusing plane located close to the tips of the micropyramids – termed the zero position in Fig. 3(b). It is illustrated by the image in a grey frame. It is repeated with the Talbot period with progressively smaller intensities, as illustrated by the grey bars in Fig. 3(b). Another subset of a half Talbot period shifted images is illustrated by the red bars in Fig. 3(b) with the representative image shown in a red frame. The peak positions are $\pi$-shifted in this subset, as it is expected for the Talbot effect. Generally, these results show that diffraction and interference grating effects are experimentally observable with Si micropyramidal arrays when illuminated with a coherent source.

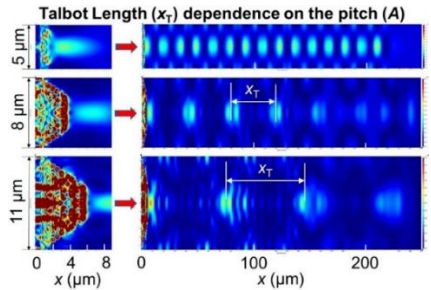

Fig. 2. EM field distributions calculated for truncated micropyramids with the size of the small base equal to 4 µm and the size of the large base equal to $A$ = 5, 8, and 11 µm, respectively. The period along $x$-axis is approximately equal to the Talbot length represented by Eq. (1).

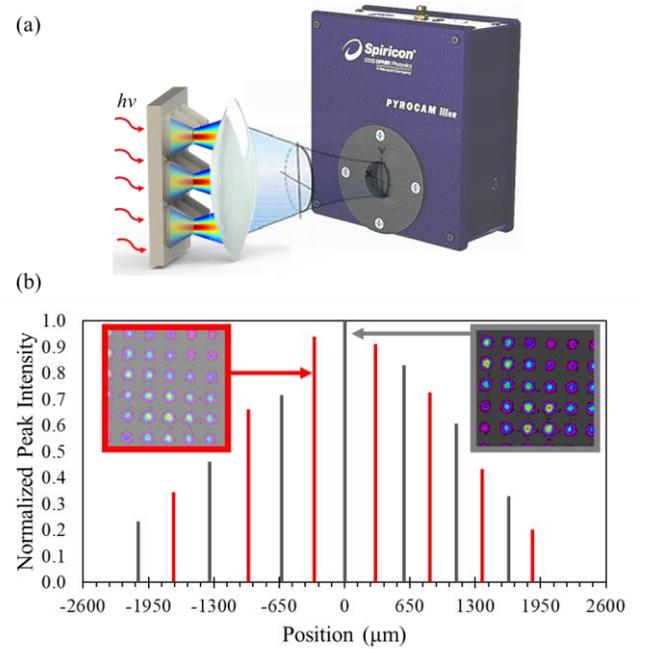

Fig. 3. (a) Experimental setup including Er:YAG laser source, Si micropyramidal array, Ge or CaF lens, and Spiricon MWIR beam profilometer. (b) Positions of the focusing planes and relative intensities of the peaks obtained from the array with the 30 µm pitch and 16 µm smaller micropyramid base are indicated by the vertical bars. There are two subsets of images where the peak positions are shifted effectively by $\pi$ as shown by the grey and red bars. For each subset the separation between the neighboring focusing planes is equal to $x_T$, whereas the shift between positions of two subsets is equal to $(1/2)x_T$.

## D. Talbot Effect: Experiment and Modeling Comparison

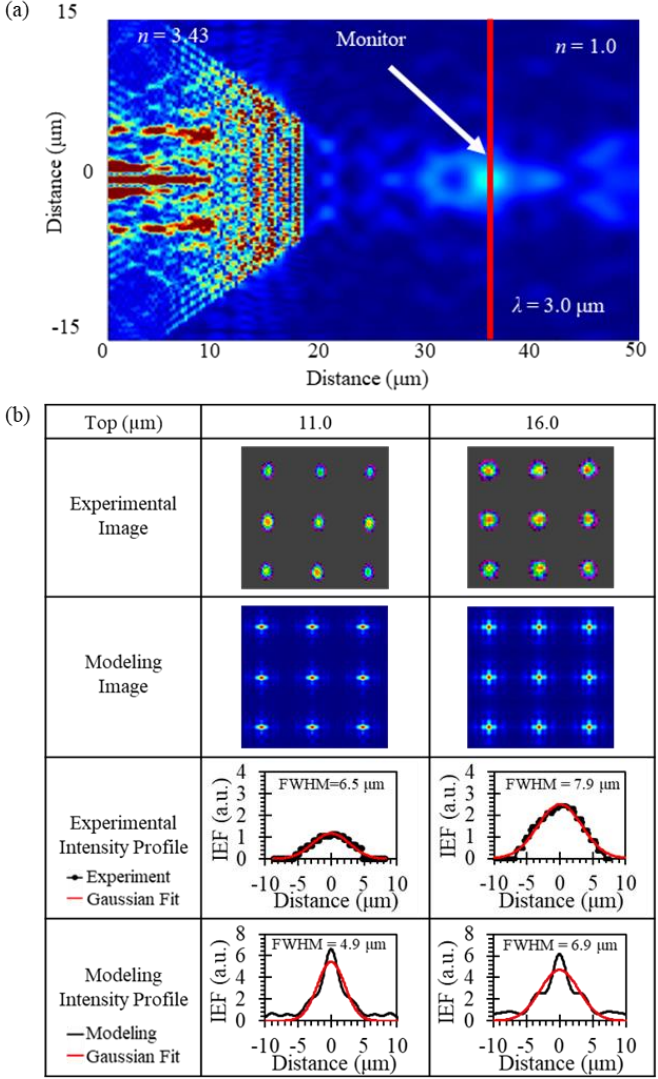

Fig. 4. (a) Electromagnetic field map for a 30.0 µm pitch with 11.0 µm top Si micropyramid when concentrating light into air. The wavelength is 3.0 µm, and the monitor position changes depending upon the micropyramid's geometry. (b) Table containing four rows consisting of experimental images, modeling images, experimental intensity profile, and modeling intensity profile for micropyramids with 11.0 µm top and 16.0 µm top micropyramids with 30.0 µm pitch. The intensity enhancement factor (IEF) is defined as the intensity in the first maximum (closest to micropyramids) of the Talbot series in interference maxima divided by the uniform reference intensity without micropyramids present. The experimental images were obtained with a Spiricon camera where the micropyramids were illuminated from the backside with a 2.96 µm wavelength Sheaumann Er:YAG laser.

In order to further demonstrate agreement between the theoretical and experimental observations of the Talbot effect, characterization of the modeling and experimental intensity distribution along the first Talbot image was performed. The first Talbot image was calculated via modeling and imaged experimentally for micropyramids with 11.0 and 16.0 µm top sizes, both with 30.0 µm pitch as illustrated in Fig. 4(b). Periodic BCs were used in the modeling. The experimental imaging setup was the same as in Fig. 3(a). A Gaussian distribution was fitted to these intensity profiles to determine their full width at half maximum (FWHM). It should be noted that the resolution limit of the experimental setup was ~ $\lambda/(2\text{NA}) = 2.1$ µm, where the numerical aperture of the lens is NA = $1/\sqrt{2}$.

With the purpose to compare the intensity distributions, the intensity enhancement factors (IEFs) were determined and can be defined as IEF = $I_{\text{pyramid}}/I_{\text{ref}}$, where $I_{\text{pyramid}}$ is the peak intensity produced by a micropyramid, and $I_{\text{ref}}$ is the uniform intensity measured without the micropyramid, as shown in Fig. 4(b). The experimental IEFs were calculated based on the experimental image obtained by the MWIR Spiricon beam profiler. The micropyramid with 11.0 µm top has a peak experimental IEF = 1.2 with FWHM = 6.5 µm, while the micropyramid with 16.0 µm top has a peak experimental IEF = 2.5 with FWHM = 7.9 µm.

Fig. 4(a) displays the results from the electromagnetic field monitor of the 11.0 µm top with 30.0 µm pitch micropyramid at $\lambda = 3.0$ µm, where the labeled power monitor spans the full 30.0 µm pitch. The power monitor's adjustable position is placed at the point of highest intensity outside the micropyramid at the first Talbot image. The modeling results show the 11.0 µm top micropyramid has a Gaussian FWHM = 4.9 µm with an IEF maximum of 6.6, while the 16.0 µm top micropyramid has a Gaussian FWHM = 6.9 µm with an IEF maximum of 6.2. The modeled IEFs of the 11.0 and 16.0 µm top micropyramids become increasingly large while simultaneously exhibiting smaller FWHMs as the pyramid top shrinks in size, consistent with the experimentally observed values shown in rows 3 and 4 of Fig. 4(b).

Therefore, the intensity distribution of the first Talbot image is found to be in a reasonable agreement regarding both their FWHMs and their IEFs. It is worth noting that the experimental values display larger FWHMs and lower IEFs compared to the modeling results, but this discrepancy can be attributed to limitations in the imaging system as well as the finite number of pixels in the MWIR Spiricon beam profiler.

## E. Light Concentrator Modeling: Perfectly Matched Layer BC

As it was discussed, the perfectly matched layer BCs remove the grating properties and instead allow us to study the light focusing properties of individual micropyramids. Most of the incident power can be delivered to the smaller base that defines power enhancement factors (PEFs) of microcones [3, 4].

In each case, the field monitor was placed at the position $x$ corresponding to the maximal intensity of the photonic nanojet easily identifiable in the calculated images on the left side of Fig. 5 due to the color bar. The dependences of the IEF and the FWHM of the photonic nanojets on the size of the smaller base, as seen in Fig. 5, show that the positions of the IEF maxima correlate with the FWHM minima. This fact is not surprising since the total photon flux proportional to (Peak IEF) × (FWHM)$^2$ is preserved along $x$. The maximal IEF~7 values can be achieved with a fairly large size of the smaller base equal to $3.7\lambda = 11.1$ µm. The photonic nanojet has FWHM~$\lambda = 3$ µm under these conditions. This is a useful result because such micropyramids are easy to fabricate in practice and, in principle, they can be integrated with various front-illuminated photodetectors.

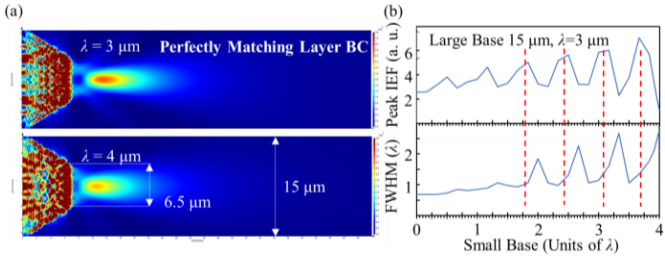

Fig. 5. (a) Images illustrating photonic nanojets calculated using perfectly matched BCs for the truncated micropyramid with 15 μm large base and 6.5 μm small base for $\lambda$=3 and 4 μm. (b) Plots showing dependences of the peak IEF and full width at the half maximum (FWHM) of photonic nanojets calculated as a function of the small base size represented in wavelength units for $\lambda$ = 3 μm.

This optimization, however, is not complete since we did not vary the pitch of the array (the size of the larger base). We plan to complete this optimization analysis in our future work. In addition, such optimization can be performed with the tips of micropyramids directly touching high index slab which would mimic the performance of the practical photodetector FPA when, for example, the photodetector material such as PbSe is deposited directly on top of the small base of micropyramids. It would be a similar situation to that considered in our recent publications [7, 8], but with the perfectly matched BC more adequately describing the focusing performance of such devices.

### III. CONCLUSION

The results of this research were three-fold:

(i) Theoretical description of the Talbot effect in micropyramidal arrays by numerical modeling with periodic BC, (ii) experimental observation of the Talbot effect in micropyramidal arrays, and (iii) estimation of the IEFs provided by micropyramids using perfectly matched layer BCs. The results demonstrate good agreement of the experimentally observed Talbot images with the theory. It is found that the photonic jets produced by the individual pyramids have typical wavelength-scale dimensions.

ACKNOWLEDGMENT

This work was supported by Center for Metamaterials, an NSF I/U CRC, award number 1068050. G.W.B. and V.N.A. received support from the AFRL Summer Faculty Fellowship Program.

REFERENCES

[1] V. N. Astratov, G. W. Bidney, J. M. Duran, G. Ariyawansa, and I. Anisimov, "Micropyramidal Photodetector Focal Plane Arrays with Enhanced Detection Capability," first draft shared between co-inventors on 08-03-2022, the provisional patent application was filed in January 2023.

[2] V. N. Astratov, A. Brettin, N.I. Limberopoulos, and A. Urbas, "Photodetector focal plane array systems and methods based on microcomponents with arbitrary shapes," US patent publication number 20190004212, 01/03/2019.

[3] B. Jin, G. W. Bidney, A. Brettin, N. I. Limberopoulos, J. M. Duran, G. Ariyawansa, I. Anisimov, A. M. Urbas, S. D. Gunapala, H. Li, and V. N. Astratov, "Microconical silicon mid-IR concentrators: Spectral, angular and polarization response," Opt. Express 28, 27615- 27627 (2020).

[4] B. Jin, A. Brettin, G. W. Bidney, N. I. Limberopoulos, J. M. Duran, G. Ariyawansa, I. Anisimov, A. M. Urbas, K. W. Allen, S. D. Gunapala, and V. N. Astratov, "Light-Harvesting Microconical Arrays for Enhancing Infrared Imaging Devices: Proposal and Demonstration," Appl. Phys. Lett. 119, 051104 (2021).

[5] M.-S. Kim, T. Scharf, C. Menzel, C. Rockstuhl, and H. P. Herzig, "Phase anomalies in Talbot light carpets of self-images," Opt. Express 21, 1287-1300 (2013).

[6] Z. Chen, A. Taflove, and V. Backman, "Photonic nanojet enhancement of backscattering of light by nanoparticles: a potential novel visible-light ultramicroscopy technique," Opt. Express 12, 1214-1220 (2004).

[7] G. W. Bidney, B. Jin, L. Deguzman, T. C. Hutchens, J. M. Duran, G. Ariyawansa, I. Anisimov, N. I. Limberopoulos, A. M. Urbas, K. W. Allen, S. D. Gunapala, and V. N. Astratov, "Fabrication of 3-D light concentrating microphotonic structures by anisotropic wet etching of silicon," Proc. SPIE 12012, Advanced FabricationTechnologies for Micro/Nano Optics and Photonics XV, 120120B (5 March 2022); doi: 10.1117/12.2610426

[8] G. W. Bidney, B. Jin, L. Deguzman, J. M. Duran, G. Ariyawansa, I. Anisimov, N. I. Limberopoulos, A. M. Urbas, K. W. Allen, S. D. Gunapala, and V. N. Astratov, "Monolithic integration of photodetector focal plane arrays with micropyramidal arrays in mid-wave infrared," Proc. SPIE 12006, Silicon Photonics XVII, 1200609 (5 March 2022); doi: 10.1117/12.2610304